\documentclass[fleqn,usenatbib]{mnras}

\usepackage{newtxtext,newtxmath}

\usepackage[T1]{fontenc}
\usepackage{lipsum}

\DeclareRobustCommand{\VAN}[3]{#2}
\let\VANthebibliography\thebibliography
\def\thebibliography{\DeclareRobustCommand{\VAN}[3]{##3}\VANthebibliography}

\usepackage{graphicx}	
\usepackage{amsmath}	

\title[High-z galactic outflows and tSZ anisotropies]{Signatures of high-redshift galactic outflows in the thermal Sunyaev Zel'dovich effect}

\author[G. Sun et al.]{
Guochao Sun$^{1}$\thanks{E-mail: guochao.sun@northwestern.edu}, Steven R. Furlanetto$^{2}$ and Adam Lidz$^{3}$
\\
$^{1}$Department of Physics, Astronomy and CIERA, Northwestern University, 1800 Sherman Ave, Evanston, IL 60201, USA\\
$^{2}$Department of Physics \& Astronomy, University of California, Los Angeles, CA 90095, USA\\
$^{3}$Department of Physics \& Astronomy, University of Pennsylvania, 209 South 33rd Street, Philadelphia, PA 19104, USA
\\
}

\date{Accepted XXX. Received YYY; in original form ZZZ}

\pubyear{2024}

\begin{document}
\label{firstpage}
\pagerange{\pageref{firstpage}--\pageref{lastpage}}
\maketitle

\begin{abstract}
Anisotropies of the Sunyaev Zel'dovich (SZ) effect serve as a powerful probe of the thermal history of the universe. At high redshift, hot galactic outflows driven by supernovae (SNe) can inject a significant amount of thermal energy into the intergalactic medium, causing a strong $y$-type distortion of the CMB spectrum through inverse Compton scattering. The resulting anisotropies of the $y$-type distortion are sensitive to key physical properties of high-$z$ galaxies pertaining to the launch of energetic SNe-driven outflows, such as the efficiency and the spatio-temporal clustering of star formation. We develop a simple analytic framework to calculate anisotropies of $y$-type distortion associated with SNe-powered outflows of galaxies at $z>6$. We show that galactic outflows are likely the dominant source of thermal energy injection, compared to contributions from reionized bubbles and gravitational heating. We further show that next-generation CMB experiments such as LiteBIRD are likely to detect the contribution to $y$ anisotropies from high-$z$ galactic outflows through the cross-correlation with surveys of Lyman-break galaxies by e.g. the Roman Space Telescope. Our analysis and forecasts demonstrate that thermal SZ anisotropies can be a promising probe of SN feedback and outflows in early star-forming galaxies. 
\end{abstract}




\section{Introduction}

The James Webb Space Telescope (JWST) has made unprecedented observations of high-redshift galaxies across different environments and mass regimes, leading to notable discoveries regarding  galaxy formation in the first billion years of cosmic time \citep{Robertson_2022,Adamo_2024}. One of the most intriguing findings about galaxy populations at $z>6$ is the prevalence of spatio-temporal clustering of star formation, namely the formation of stars in spatially clustered clumps \citep[e.g.][]{Fujimoto_2024,Mowla_2024} and with a temporally stochastic (or `bursty') star formation history \citep[e.g.][]{Ciesla_2024,Dressler_2024,Endsley_2024}. Interestingly, high-resolution cosmological simulations of galaxy formation suggest a similar picture that early galaxies are characterized by spatially and temporally clustered star formation \citep[e.g.][]{Ma_2018, Sun_2023a, Sun_2023b, Bhagwat_2024}. 

Stellar feedback from supernovae (SNe) is considered as a main physical driver of the spatio-temporal clustering of star formation in galaxies \citep{Hopkins_2023,Hu_2023}. Besides regulating star formation, SN feedback also drives multi-phase outflows with particularly strong rates at high redshift where galaxies are more gas rich and less massive \citep{HaywardHopkins_2017,Furlanetto_2021}. An essential part of the cosmic baryon cycle, galactic outflows inform the efficiency at which SN feedback regulates stellar and gas mass buildup in galaxies while depositing energy and metal-enriched gas into the intergalactic medium (IGM). A complete picture of high-$z$ galaxy formation thus requires understanding galactic outflows. 

The cosmic microwave background (CMB) is a powerful and versatile tool to probe the collective effects of early galaxy populations in a highly complementary way to galaxy observations. Its polarization provides an integral constraint on the history of cosmic reionization driven by the UV radiation from galaxies \citep{Miranda_2017,Pagano_2020}. High-$z$ galaxies and reionization also leave measurable imprints on the CMB spectrum through the Sunyaev-Zel'dovich (SZ) effect \citep{SZ_1980} associated with the inverse Compton scattering of CMB photons on free electrons. The bulk motion of electrons created during and after patchy reionization causes the kinetic SZ (kSZ) effect, a Doppler shift whose spatial fluctuations constrain the reionization timeline and morphology \citep{Battaglia2013,Gorce_2020,Chen_2023,Jain_2024}. The thermal motion of electrons, on the other hand, causes the thermal SZ (tSZ) effect characterized by the Compton $y$ parameter. This $y$-type distortion of the CMB spectrum is a promising probe of the thermal energy of gas surrounding early galaxies due to both reionization and winds launched by SNe \citep{Oh_2003,Baxter_2021,Namikawa_2021,Yamaguchi_2023}. 

In this Letter, motivated by the recent progress in understanding high-$z$ galaxies and the planned next-generation CMB experiments like LiteBIRD \citep{LiteBIRD_Collaboration_2023}, we analytically estimate the $y$-type distortion and its spatial fluctuations (i.e. anisotropies) induced by the thermal energy injection from high-$z$ galaxies especially their SNe-driven winds. We then compare the derived anisotropies with the expected CMB measurement uncertainties and its potential synergy with galaxy surveys to estimate the constraining power on $y$. Throughout, we assume a Chabrier initial mass function \cite[IMF;][]{Chabrier_2003} and a flat, $\Lambda$CDM cosmology consistent with measurements by \citet{Planck_2016}. 

\section{Models}

\subsection{Cooling of SNe-powered galactic wind bubbles} \label{sec:models:winds}

Recent observations and simulations have suggested that star formation might be highly clumpy and bursty in high-$z$ galaxies. Therefore, rather than considering supernova explosions randomly distributed in the ISM, we consider the possibility that clustered SNe can drive superbubbles by launching more powerful galactic winds into the IGM \cite[see e.g.][and references therein]{Fielding_2017,Fielding_2018}. The breakout of superbubbles driven by clustered SNe minimizes radiative losses in the interstellar medium (ISM), thus depositing a substantially higher fraction of SNe energy into the IGM, which can then be transferred to the CMB via Compton cooling.  

To roughly estimate the fractions of energy released by supernova explosions lost through radiative and inverse Compton processes, we follow \citet{Oh_2003} and compare the timescales of isobaric radiative cooling, $t_\mathrm{rad}$, to the ambient gas, and of inverse Compton cooling, $t_\mathrm{comp}$, to CMB photons in the IGM. The fraction of explosion energy deposited in the IGM which thus modifies the CMB spectrum can be approximated as
\begin{equation}
\varepsilon \approx \frac{1/t_\mathrm{comp}}{1/t_\mathrm{comp} + 1/t_\mathrm{rad}},
\label{eq:epsilon}
\end{equation}
with
\begin{equation}
t_\mathrm{comp} = \frac{3 m_e c}{8 \sigma_\mathrm{T} a T_\mathrm{CMB}^4} = 5\times10^8 \left(\frac{1+z}{7}\right)^{-4}\,\mathrm{yr},
\end{equation}
where $\sigma_\mathrm{T}$ is the Thomson scattering cross section for electrons and $a=7.57\times10^{-15}\,\mathrm{erg\,cm^{-3}\,K^{-4}}$, and
\begin{equation}
t_\mathrm{rad} = \frac{3 k_\mathrm{B} T}{2 n \Lambda(T)},
\end{equation}
where $\Lambda(T)$ is the cooling function at the postshock temperature $T$ and $n$ is the number density of the ambient gas. Within the inhomogeneous ISM, the cooling due to radiative losses can be expressed as \citep{Martizzi_2015,Fielding_2018}
\begin{equation}
t_\mathrm{rad} = 5 \times 10^{3} \left(\frac{n}{100\,\mathrm{cm^{-3}}} \right)^{-0.53} \left( \frac{Z}{0.2\,Z_{\odot}} \right)^{-0.17} \,\mathrm{yr}, 
\end{equation}
which is a few orders of magnitude shorter than $t_\mathrm{comp}$ even if large density fluctuations exist in the turbulent ISM.  

However, considering the situation where the expansion and eventual breakout of winds from the galactic disc \citep{KooMcKee_1992,Cooper_2008} are powered by clustered SNe, \citet{Fielding_2018} suggest that SNe exploded in massive star clusters efficiently overlap in space and time to collectively power superbubbles \cite[see also][]{Kim_2017}. The cooling time for radiative losses is significantly longer than the expansion time $t_\mathrm{exp} = R_\mathrm{s}/v_\mathrm{s}$ of bubbles out to scales of $\sim$100\,pc if the mixing of SNe ejecta and the ambient ISM remains inefficient
\begin{equation}
\frac{t_\mathrm{rad}}{t_\mathrm{exp}} \simeq 100 \left( \frac{n}{100\,\mathrm{cm^{-3}}} \right)^{-2/3} \left( \frac{M_\mathrm{cl}}{10^{5}M_{\odot}} \right)^{-1/3} \left( \frac{R_\mathrm{s}}{100\, \mathrm{pc}} \right)^{-1/3}. 
\label{eq:t_ratio_pre}
\end{equation}
Thus, radiative cooling can be very modest throughout the bubble expansion within galaxies even if the gas is as dense as $n\sim10^3\,\mathrm{cm^{-3}}$, a typical threshold for star formation. Although models of disc galaxy evolution may not be entirely valid at $z>6$, $R_\mathrm{s}$ of order 100\,pc is still a relevant scale for the breakout of SNe-powered superbubbles into ambient gas with significantly lower densities, given the typical sizes of high-$z$ galaxies \citep{Morishita_2024} and spatially-resolved star-forming clumps \citep{Fujimoto_2024} observed by the JWST. In low-mass galaxies, SNe feedback may be strong enough to rapidly blow out the dense gas entirely, making radiative losses before bubble breakout even less of an effect. 

\begin{figure}
	\includegraphics[width=\columnwidth]{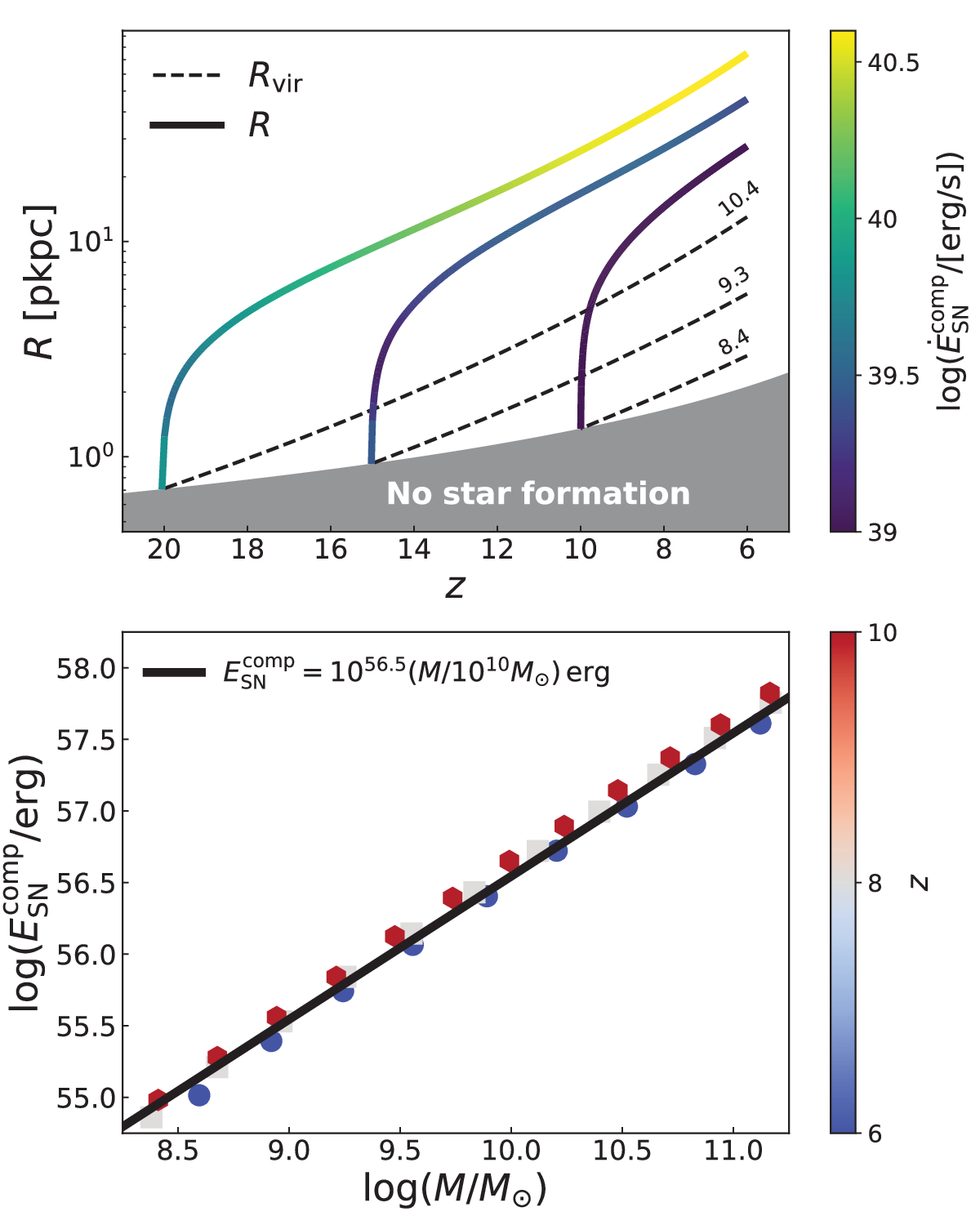}
    \vspace{-1em}
    \caption{\textit{Top:} the expansion of galactic wind-driven bubbles (measured in bubble radius $R$) compared to the growth of halo virial radius $R_\mathrm{vir}$. The 3 example haloes begin to evolve from $z = 20, 15$, and 10 with $\log(M_\mathrm{min}/M_{\odot}) = 8$ (below which stars do not form) and stop at $z=6$, reaching a mass of $\log(M/M_{\odot}) = 8.4, 9.3, 10.4$, respectively. The colour coding indicates the rate of energy loss via Compton cooling, whose time integral $E^\mathrm{comp}_\mathrm{SN}$ measures the contribution to the tSZ effect from SNe-powered winds. \textit{Bottom:} $E^\mathrm{comp}_\mathrm{SN}$ as a function of halo mass derived from the bubble expansion model at $z=6$ (blue circles), $8$ (grey squares), and $10$ (red hexagons). Results in our fiducial case are well fit by a simple, redshift-independent power law: $E^\mathrm{comp}_\mathrm{SN} = 3 \times 10^{56}(M_\mathrm{h}/10^{10}M_{\odot})\,\mathrm{erg}$.}
    \label{fig:bubble}
\end{figure}

Following \citet{Yamaguchi_2023}, we can model the wind bubble expansion after breakout by solving a system of differential equations for the bubble radius $R$, interior pressure $P$, and shell mass $M_\mathrm{s}$ under the thin-shell approximation \citep[see also][]{Tegmark_1993,FurlanettoLoeb_2003}: 
\begin{equation}
\ddot{R} = \frac{4\pi R^2}{M_\mathrm{s}} P - \frac{G}{R^2}\left(M + \frac{4\pi}{3} \bar{\rho}_{M} R^3 + \frac{M_\mathrm{s}}{2}\right) - \frac{\dot{M}_\mathrm{s}}{M_\mathrm{s}}\left(\dot{R} - HR\right),
\end{equation}
\begin{equation}
\dot{P} = \frac{\dot{E}_\mathrm{SN} + \dot{E}^\mathrm{cool}_\mathrm{SN}}{2\pi R^3} - 5 P \frac{\dot{R}}{R},
\end{equation}
\begin{equation}
\dot{M}_\mathrm{s} = 4\pi R^2 \bar{\rho}_\mathrm{b} (\dot{R}-HR) \mathcal{H}(\dot{R}-HR),
\end{equation}
where $\bar{\rho}_{M}$ and $\bar{\rho}_\mathrm{b}$ are the mean dark matter and baryon densities and $\mathcal{H}$ is the unit step function that sets the bubble comoving size fixed once $\dot{R} < HR$. The boundary conditions are set such that the wind launches at the virial radius, $R_\mathrm{vir}$, once a halo begins forming stars ($M_\mathrm{h} > 10^8 M_{\odot}$). The energy injection rate from SNe
\begin{equation}
\dot{E}_\mathrm{SN} = f_{*}(\Omega_\mathrm{b}/\Omega_\mathrm{m})\dot{M}\varepsilon_{0}\omega_\mathrm{SN},
\end{equation}
where $\dot{M}$ is the halo mass accretion rate that can be determined from abundance matching \citep{Furlanetto_2017}. While equation~(\ref{eq:t_ratio_pre}) suggests radiative losses in the ISM can be small due to clustered SNe, for the fiducial case, we conservatively assume $\varepsilon_0 = 0.3$ as the fraction of SN energy available for driving winds after bubble breakout \citep{Fielding_2018}. We assume a star formation efficiency (SFE) $f_{*} = 0.1$ and $\omega_\mathrm{SN} = 2 \times 10^{49}\,$erg ($10^{51}\,$erg per 50$M_{\odot}$ of stellar mass formed) for a Chabrier IMF \citep{Cen_2020}. The cooling rate
\begin{equation}
\dot{E}^\mathrm{cool}_\mathrm{SN} = -(3/2)PV/t_\mathrm{eff} = -2\pi R^3/t_\mathrm{eff},
\end{equation}
where $t^{-1}_\mathrm{eff} = t^{-1}_\mathrm{rad} + t^{-1}_\mathrm{comp}$. We evaluate $t_\mathrm{rad}$ for a radiative cooling rate of $\Lambda(T) = 3\times10^{-24}\,\mathrm{erg\,s^{-1}\,cm^{3}}$ and a number density of $n \sim 4 n_\mathrm{IGM} \sim 10^{-3}[(1+z)/7]^{3}\,\mathrm{cm^{-3}}$, appropriate for the intergalactic gas with primordial composition at temperature $T\sim10^5$--$10^6\,$K since hot gas dominates the energy loading of outflows \citep{Pandya_2021}. Note that radiative cooling post-breakout can be more efficient if wind bubbles start at $R \ll R_\mathrm{vir}$ and expand into denser circumgalactic gas. Such uncertainties can be partially absorbed in the assumed value of $\varepsilon_{0}$, which we vary when characterizing the tSZ signal (see Section~\ref{sec:results}). More detailed modeling of the bubble expansion, including radiative losses at different stages, is postponed to future studies.

We show in Figure~\ref{fig:bubble} the solution to the bubble expansion model in our fiducial case, together with the implied thermal energy loss due to Compton cooling. The bottom panel, in particular, suggests that the integrated thermal energy contribution to the CMB from SNe-powered winds via Compton cooling can be well approximated by a simple (redshift-independent) power law in halo mass.

\subsection{CMB spectral distortions induced by high-redshift galaxies}

The Compton-$y$ parameter that characterizes the tSZ effect directly probes the line-of-sight integral of the proper electron gas pressure, $P_\mathrm{e}$ (or equivalently the electron energy density), namely
\begin{equation}
y = \frac{\sigma_\mathrm{T}}{m_\mathrm{e} c^2} \int \mathrm d z \frac{1}{1+z} \frac{\mathrm d \chi}{\mathrm d z} P_\mathrm{e}(z) = \frac{\sigma_\mathrm{T}}{m_\mathrm{e} c} \int \mathrm d z \frac{P_\mathrm{e}(z)}{(1+z)H(z)},
\label{eq:y_from_pe}
\end{equation}
where $m_\mathrm{e}$ is the electron mass and $\chi$ is the comoving radial distance. 

In addition to distortions in the sky-averaged CMB spectrum, equation~(\ref{eq:y_from_pe}) also implies a large-scale clustering signal of $y$-type distortion anisotropies that scales as
$C_{\ell, \mathrm{2h}} \propto y^2$. 
It is therefore instructive to compare thermal energy contributions to $\langle b P_\mathrm{e} \rangle$ from different sources, including SNe-powered galactic winds, reionized bubbles, and gravitational heating. Considering the halo model \citep{CS_2002}, we can sum up the thermal energy contributed from individual haloes to obtain the average bias-weighted (proper) electron pressure \citep{VLJ_2017}
\begin{equation}
\langle b P_\mathrm{e} \rangle = \int \mathrm d M \frac{\mathrm d n}{\mathrm d M} \left( E^\mathrm{comp}_\mathrm{SN} + E_\mathrm{reion} + E_\mathrm{G} \right) (1+z)^3 b(M),
\label{eq:energy_decompose}
\end{equation}
where $\mathrm d n / \mathrm d M$ is the halo mass function and $b(M)$ is the halo bias. 

Taking results from Section~\ref{sec:models:winds} and the best-fit power law in Figure~\ref{fig:bubble}, we can approximate the thermal energy injected by the SNe-powered galactic wind bubbles (and available for Compton cooling by CMB photons) as
\begin{equation}
E^\mathrm{comp}_\mathrm{SN} \approx 3\times10^{56}\left(\frac{M}{10^{10}M_{\odot}}\right) \left(\frac{ \omega_\mathrm{SN}}{2\times10^{49}\mathrm{erg}}\right) \left( \frac{f_*}{0.1} \right) \left(\frac{\varepsilon_0}{0.3}\right)\,\mathrm{erg}, 
\label{eq:E_SN}
\end{equation}
where we assume $f_{*} \varepsilon_0 = 0.03$ in our fiducial case. This corresponds to an effective (time-integrated) Compton efficiency of about 10\%. The thermal energy of reionized bubbles can be estimated by
\begin{equation}
E_\mathrm{reion} = \frac{A_\mathrm{He} \zeta \Omega_\mathrm{b} M \mathcal{E}}{\Omega_\mathrm{m} m_\mathrm{p}} \approx 5\times10^{55} \left(\frac{M}{10^{10}M_{\odot}}\right)\,\mathrm{erg},
\label{eq:E_reion}
\end{equation}
where $A_\mathrm{He}=1.22$, $\zeta = f_* f_\mathrm{esc} N_{\gamma}$ with $f_\mathrm{esc} = 0.1$ and $N_{\gamma}=4000$, and $\mathcal{E}=1\,\mathrm{eV}$ is the typical energy of the reionized medium photo-heated to roughly $10^4\,$K. With $f_*$ canceled out, the ratio $E^\mathrm{comp}_\mathrm{SN}/E_\mathrm{reion}$ thus scales as $\varepsilon_{0} f^{-1}_\mathrm{esc}$. Note that equation~(\ref{eq:E_reion}) assumes that the ionized bubble is created for the first time, ignoring previous photoheating. Thus, by $z\sim6$, it in fact serves as an upper limit on the thermal energy due to reionization. Finally, as an upper limit on the thermal energy contributed by gravitational heating of gas inflows, we have
\begin{equation}
E_\mathrm{G} = \frac{3k_\mathrm{B}T_\mathrm{vir}}{2} \frac{\Omega_\mathrm{b} M}{\Omega_\mathrm{m} \mu m_\mathrm{p}} \approx 10^{56} \left(\frac{M}{10^{10}M_{\odot}}\right)^{5/3} \left(\frac{1+z}{7}\right)\,\mathrm{erg}.
\label{eq:E_G}
\end{equation} 
In practice, though, efficient radiative cooling can prevent gas from being shocked-heated to the virial temperature $T_\mathrm{vir}$ by gravity. For reference, hot gas with $n = 50\,n_\mathrm{IGM} \sim 10^{-2}\,\mathrm{cm^{-3}}$ at $z=6$ can have strong radiative losses since $t_\mathrm{rad}/t_\mathrm{comp} \lesssim 0.2$.

\subsection{Auto-/cross-correlations of $y$-type distortion anisotropies} \label{sec:models:statistics}

The $y$-type spectral distortion power spectrum from the tSZ effect associated with high-$z$ galaxies can be written in the large-scale, two-halo limit\footnote{We note that a non-negligible one-halo term may arise from extended/overlapping outflows and ionized bubbles. However, since observational constraints considered here are most sensitive to $\ell \lesssim 1000$ (equivalent to physical scales $\gtrsim3\,$Mpc at $z=6$), we ignore its contribution in our analysis.} as
\begin{equation}
C^{yy}_{\ell} = \int \mathrm d z \frac{H(z)}{c \chi^2} P_\mathrm{m} \left( \frac{\ell}{\chi} \right) \left[ \frac{\sigma_\mathrm{T}}{m_\mathrm{e} c^2} \frac{1}{1+z} \frac{\mathrm d \chi}{\mathrm d z} \langle b(z) P_\mathrm{e}(z) \rangle \right]^2,
\end{equation}
which is approximately proportional to $\bar{y}^2$ as is evident from equations~(\ref{eq:y_from_pe}) and (\ref{eq:energy_decompose}). 

\begin{figure*}
	\includegraphics[width=\columnwidth]{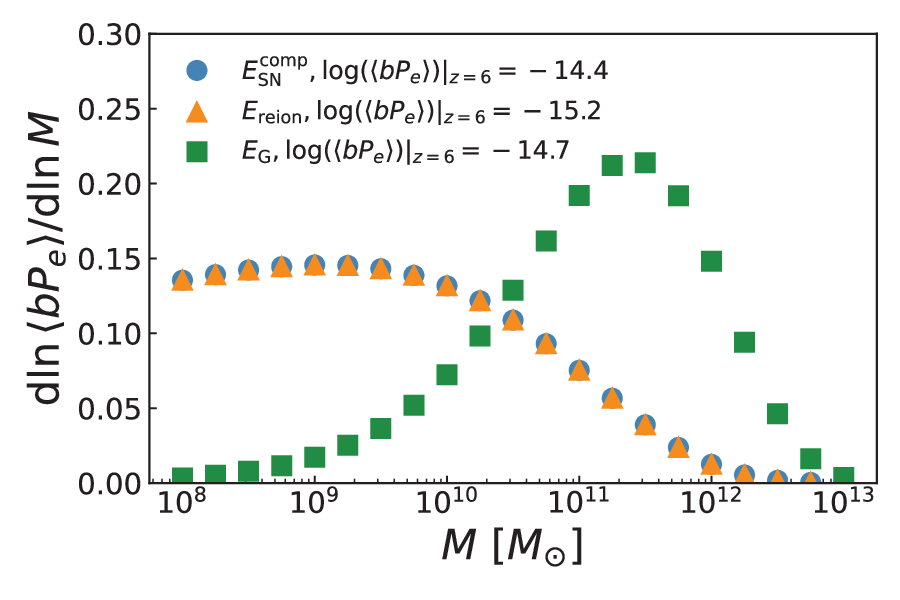}
	\includegraphics[width=\columnwidth]{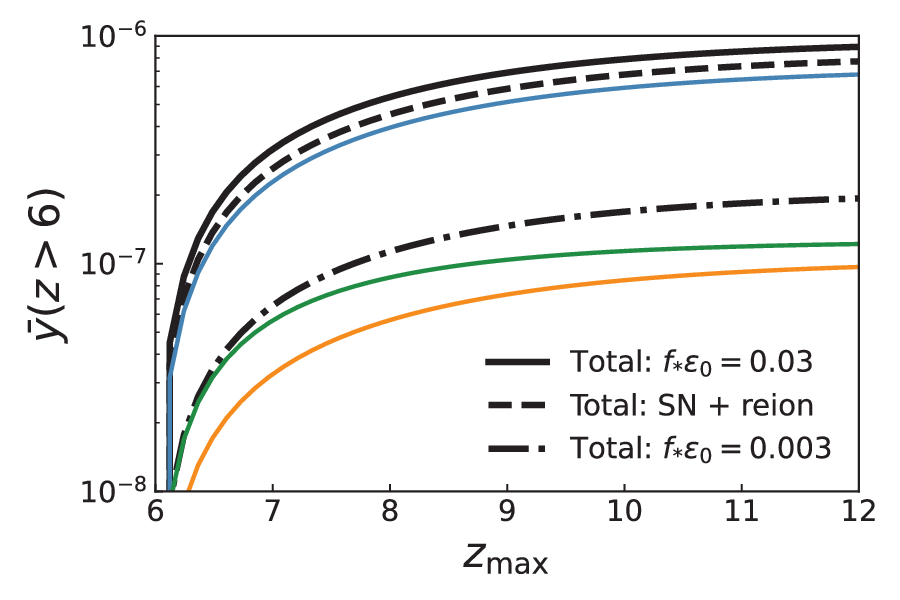}
    \vspace{-1em}
    \caption{Comparison of the average thermal energy available and $\bar{y}$ contributions from SNe-driven winds, reionized bubbles, and gravitational heating. \textit{Left:} fractional contributions to $\langle b P_\mathrm{e} \rangle$ from haloes of different masses at $z=6$. The exact value of $\langle b P_\mathrm{e} \rangle$ at $z=6$ (in units of $\mathrm{erg\,cm^{-3}}$) from each source of energy is labeled. \textit{Right:} the average Compton-$y$ parameter associated with all haloes with $M > 10^8\,M_{\odot}$ between $z_\mathrm{min}=6$ and $z_\mathrm{max}$, evaluated with equation~(\ref{eq:y_from_pe}). The black curves show the total $\bar{y}$ signal in the fiducial case with $f_* \varepsilon_0 = 0.03$ (solid), neglecting $E_\mathrm{G}$ (dashed), and with $f_* \varepsilon_0 = 0.003$ (dash-dotted), respectively. The 3 coloured curves indicate the contributions from SNe, reionization, and gravitational heating in the fiducial case.}
    \label{fig:bPe}
\end{figure*}

The cross-correlation of $y$-type distortion anisotropies with galaxy surveys provides an alternative and likely more reliable way to constrain the energy of SNe-powered high-$z$ galactic winds, given the significant low-$z$ contributions in $y$ measurements that are challenging to remove. The $y$--galaxy cross-correlation is advantageous because low-$z$ contributions to $y$ only add to the variance but do not bias the measurement. On large scales, the cross-power spectrum is
\begin{align}
C^{y\mathrm{g}}_{\ell} & = \int \frac{\mathrm d z H(z)}{c \chi^2(z)} P_\mathrm{m} \left[ \frac{\ell}{\chi(z)}, z \right] f_\mathrm{g}(z) f_{y}(z) b_\mathrm{g}(z) \langle b(z) P_\mathrm{e}(z) \rangle
\end{align}
where the $y$ window function $f_{y}(z) = (\mathrm d \chi / \mathrm d z) \sigma_\mathrm{T} / (m_\mathrm{e} c^2) / (1+z) $ and the top-hat galaxy window function $f_\mathrm{g}(z)$ is $1/\Delta z_\mathrm{g}$ over $z_\mathrm{g} \pm \Delta z_\mathrm{g}/2$ and zero elsewhere. 

The uncertainties of the auto- and cross-power spectra can be expressed as \citep{Baxter_2021}
\begin{equation}
\Delta C^{yy}_{\ell} = \frac{1}{\sqrt{f_\mathrm{sky} (\ell+1/2) \Delta\ell}} \left[ C^{yy}_{\ell} + C^{yy}_{\ell,\mathrm{N}} \right]
\label{eq:cyy_unc}
\end{equation}
and
\begin{equation}
\Delta C^{y\mathrm{g}}_{\ell} = \sqrt{\frac{\left( C^{yy}_{\ell} + C^{yy}_{\ell,\mathrm{N}} \right)\left( C^\mathrm{gg}_{\ell} + C^\mathrm{gg}_{\ell,\mathrm{N}} \right)+ \left(C^{y\mathrm{g}}_{\ell}\right)^2}{f_\mathrm{sky} (2\ell+1) \Delta \ell}},
\label{eq:cyg_unc}
\end{equation}
where $f_\mathrm{sky}$ is the sky covering fraction and $\Delta \ell$ is the multipole bin width centered at $\ell$. The terms $C^{yy}_{\ell,\mathrm{N}}$, $C^\mathrm{gg}_{\ell}$, $C^\mathrm{gg}_{\ell,\mathrm{N}}$ are the instrument noise power spectrum for the $y$-type distortion, the galaxy angular power spectrum, and the galaxy noise power spectrum (equal to the inverse of galaxy number density), respectively. 

\section{Results} \label{sec:results}

\subsection{Contributions to $y$ from different sources of thermal energy}

We show in the left panel of Figure~\ref{fig:bPe} the fractional contribution of each source of thermal energy to $\langle b P_\mathrm{e} \rangle$ at $z = 6$ for different halo masses. Because of the identical linear dependence on $M$, the SNe and reionization components have the same larger contributions from low-mass haloes with $M < 10^{10}\,M_{\odot}$, whereas haloes with $M \sim 10^{11.5}\,M_{\odot}$ contribute most to the gravitational heating term given its stronger $M$ dependence. Note again, though, that equation~(\ref{eq:E_G}) may significantly overestimate the true $E_\mathrm{G}$ due to radiative cooling. 

The right panel of Figure~\ref{fig:bPe} shows the level of $\bar{y}$ signal contributed by haloes at different redshifts, as predicted by our fiducial model, along with an alternative scenario where $f_{*} \varepsilon_0 = 0.003$ that accounts for model uncertainties in both $f_{*}$ and $\varepsilon_0$. Given our fiducial parameters, we expect $\bar{y}$ associated with galaxies above $z = 6$ to be as large as $10^{-6}$, with a significant ($>$$50\%$) contribution from SNe-driven galactic outflows. Even with $f_{*} \varepsilon_0 = 0.003$ that can result from a lower SFE or less energetic winds due to e.g. stronger radiative losses before and/or after bubble breakout, the redshift-integrated $\bar{y} > 10^{-7}$ (see Section~\ref{sec:discussion} for more discussion). Notably, these predicted $\bar{y}$ values are consistent with the observational constraints available from COBE FIRAS \citep[$\bar{y} < 1.5\times10^{-5}$;][]{Fixsen_1996} and a joint analysis of Planck and SPT data \citep[$5.4\times10^{-8}<\bar{y}<2.2\times10^{-6}$;][]{KhatriSunyaev_2015}. As will be shown below, the LiteBIRD satellite to be launched in 2032 is expected to substantially increase the detectability of $\bar{y}$ from large-scale tSZ anisotropies \citep{Miyamoto_2014,Remazeilles_2024} and place useful constraints on early galaxy formation through its high-$z$ contribution. 

\subsection{Anisotropies of the $y$-type distortion from high-$z$ galaxies}

With the estimated $\bar{y}$ associated with high-$z$ galaxies in hand, we can calculate the anisotropy signals and compare them against the sensitivity level of CMB observations as described in Section~\ref{sec:models:statistics}. In contrast with previous studies, we find $\bar{y}$ and $C^{yy}_{\ell}$ to be slightly below even the most pessimistic model of \cite{Oh_2003} in which a higher $\varepsilon$ is assumed. Our predicted $\bar{y}$, however, is systematically larger than those ($\bar{y} \approx 10^{-8}$--$10^{-7}$) from \citet{Hill_2015} and \citet{Yamaguchi_2023} as a combined result of the (effectively) larger $\varepsilon_0$ and $f_*$ in our fiducial model, which are nevertheless motivated by our current understanding of high-$z$ galaxy populations even though some caveats apply (see Section~\ref{sec:discussion}).

We can compare our predicted $C^{yy}_{\ell}$ signal against the sensitivity levels of forthcoming CMB experiments like LiteBIRD that promise to make precise measurements of tSZ anisotropies. Following \citet{Miyamoto_2014}, we take the instrument noise power of LiteBIRD to be $C_{\ell,\mathrm{N}}^{yy} = 3.3\times10^{-19} e^{(\ell/135)^2}+2.8\times10^{-19} e^{(\ell/226)^2}$. On degree scales ($\ell \approx 100$), $C^{yy}_{\ell}$ of $z>6$ galaxies lies above the 1-$\sigma$ uncertainty level of LiteBIRD. Considering $50<\ell<150$ and an $f_\mathrm{sky}$ close to unity, we can use equation~(\ref{eq:cyy_unc}) to estimate the signal-to-noise ratio (S/N) of the pure $C^{yy}_{\ell}$ signal to be greater than 3. However, in reality, it would be challenging to separate or remove the low-$z$ contribution to $\bar{y}$ (by methods like masking, see e.g. \citealt{Baxter_2021}), whose residuals can easily overwhelm the high-$z$ signal of interest \citep{Oh_2003}. It is thus instructive to consider the $y$--galaxy cross-correlation that allows a cleaner and unbiased measurement of $\bar{y}$ truly associated with high-$z$ galaxies \citep{Baxter_2021}.

\subsection{Detectability of the $y$--galaxy cross-correlation} \label{sec:results:cyg}

The full-sky survey strategy of LiteBIRD makes it convenient to cross-correlate with other cosmological data sets, as discussed in several previous studies \citep{Namikawa_2023,Lonappan_2024}. To assess whether $y$-type distortion anisotropies of high-$z$ origin can be detected when residual low-$z$ contributions are present, we consider a case study for a joint analysis between LiteBIRD and the High Latitude Wide Area Survey (HLWAS) of the Nancy Grace Roman Space Telescope. Assuming $f_\mathrm{sky} = 0.05$ corresponding to the planned 2200\,deg$^2$ coverage of HLWAS and a 5-sigma depth of $m_\mathrm{AB} = 26.5$, we adopt the galaxy number density and bias estimated for HLWAS by \citet{LaPlante_2023} to evaluate $C^{y\mathrm{g}}_{\ell}$, $C^\mathrm{gg}_{\ell}$, and $C^\mathrm{gg}_{\ell,\mathrm{N}}$ in the redshift range of interest. 

\begin{figure}
    \centering
	\includegraphics[width=\columnwidth]{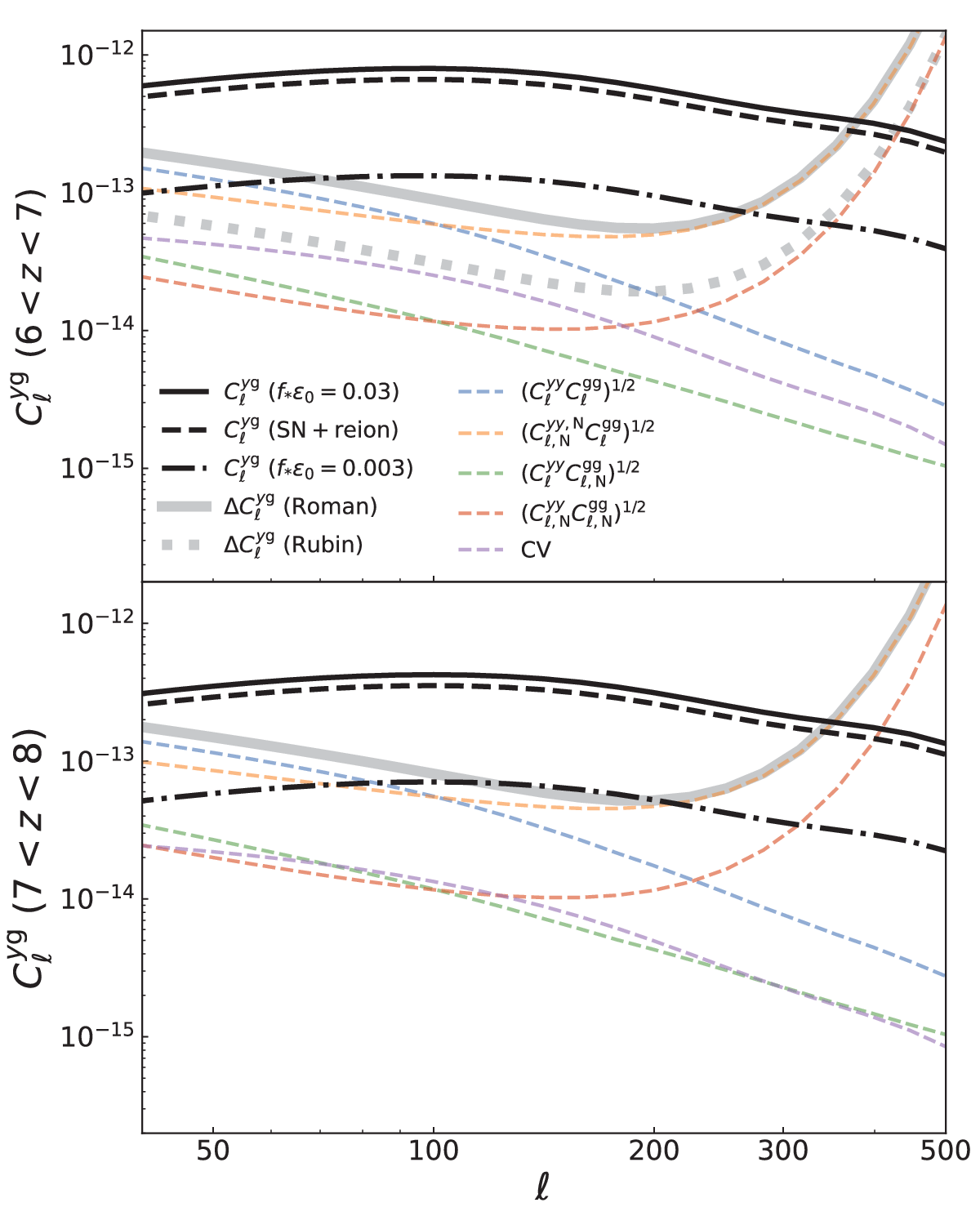}
    \vspace{-1em}
    \caption{The angular cross-power spectrum $C^{y\mathrm{g}}_{\ell}$ of the Compton-$y$ map to be measured by LiteBIRD and LBGs to be observed by the Roman High Latitude Wide Area Survey over $6 < z < 7$ (top) and $7 < z < 8$ (bottom). For comparison, a factorization of the cross-power spectrum uncertainty as specified in equation~(\ref{eq:cyg_unc}) is also shown for our fiducial case, which implies that the uncertainty is mainly dominated by the residual low-$z$ contribution to $\bar{y}$ and the instrument noise of LiteBIRD rather than the cosmic variance. We estimate the total S/N of $C^{y\mathrm{g}}_{\ell}$ for LiteBIRD and Roman to be approximately 10 and 5 for the top and bottom panels, respectively. In the top panel, we further show the estimated uncertainty level for LBGs observable by Rubin/LSST up to $z\sim7$, which is roughly 3 times smaller thanks to a larger $f_\mathrm{sky}$. }
    \label{fig:cygal}
\end{figure}

Figure~\ref{fig:cygal} shows the cross-power spectrum $C^{y\mathrm{g}}_{\ell}$ between the $y$-type distortion to be measured by LiteBIRD and Lyman-break galaxies (LBGs) to be observed by the Roman HLWAS survey, together with a decomposition of the uncertainty $\Delta C^{y\mathrm{g}}_{\ell}$ as given by equation~(\ref{eq:cyg_unc}). Note that here we have included a low-$z$ contribution (after a reasonable level of masking) to the auto-power term $C^{yy}_{\ell}$ using the estimate in \citet{Baxter_2021}, which is roughly two orders of magnitude stronger than the high-$z$ contribution in our fiducial model. We have also assumed that the distinctive y-distortion spectral signature allows perfect component separation. That is, we assume negligible residual foreground contamination in the multi-frequency LiteBIRD data from cosmic infrared background fluctuations and other components \citep{Remazeilles_2024}.

The comparison suggests that even with the inclusion of a strong residual low-$z$ contribution to $\bar{y}$, thanks to the wide overlapping area and the large number of galaxies available the $y$--galaxy cross-correlation is likely detectable on scales $\ell \sim 150$. More specifically, summing over the $\ell$ bins (with $\Delta \ell \approx \ell$) over $50 < \ell < 300$, we estimate the total S/N of $C^{y\mathrm{g}}_{\ell}$ in our fiducial case with $f_{*} \varepsilon_0 = 0.03$ to be about 10 and 5 for the two bins over $6<z<7$ and $7<z<8$, respectively. Even for the more pessimistic scenario with $f_{*} \varepsilon_0 = 0.003$, the cross-correlation may still be marginally detectable over $6<z<7$. These results imply that measurements of $C^{y\mathrm{g}}_{\ell}$ from a synergy of LiteBIRD and Roman can potentially constrain or at least perform null tests of the $y$-type distortion associated with high-$z$ galaxies, in particular the production of their SNe-driven outflows that dominate the thermal energy content (see Figure~\ref{fig:bPe}). 

In addition to Roman, we show over $6<z<7$ the improved detectability of $C^{y\mathrm{g}}_{\ell}$ from accessing a larger $f_\mathrm{sky}$ by cross-correlating with LBGs from the Rubin Observatory Legacy Survey of Space and Time (LSST; \citealt{Ivezic_2019}). Assuming Rubin/LSST LBG samples at the same depth as Roman but with $f_\mathrm{sky} = 0.44$, we expect a roughly threefold increase in $C^{y\mathrm{g}}_{\ell}$ sensitivity. This adds to both the $\ell$ range probed and the constraining power for a wider range of parameter combinations, especially less optimistic ones. For example, models with $f_{*} \varepsilon_0$ as small as 0.001 may still be meaningfully constrained up to $z\sim7$ by LiteBIRD and Rubin/LSST. 

\section{Discussion} \label{sec:discussion}

It is worth noting that the roughly comparable thermal energy contributions from SNe-driven winds and reionized bubbles we find should be contrasted with previous simulation-based analyses such as \citet{Baxter_2021} with caution for two main reasons. First, the simulations do not distinguish thermal energies attributed to galactic winds and the ionized bubble created by the galaxy, but rather focus on the total thermal energy content. Second, the weak dependence of $y$ signal on the prescription of stellar feedback reported in \citet{Baxter_2021}, which might be interpreted as evidence for a subdominant role of SNe-driven winds, can be (at least) partially due to the limited resolution of the simulations to properly resolve the clustering of SNe. As discussed in Section~\ref{sec:models:winds}, superbubbles driven by clustered SNe can be the main reason for a significant fraction of SNe energy to be vented into the IGM and power wind bubbles. This too explains the overall larger $y$ anisotropies predicted by our analytic calculations. 

Several aspects of our simple model should be further investigated in the future for better understanding the tSZ signal associated with high-$z$ galaxies. First, as demonstrated by the comparison of our fiducial ($f_* \varepsilon_0 = 0.03$) and more pessimistic ($f_* \varepsilon_0 = 0.003$) cases, the amplitude of the thermal energy associated with SNe-powered winds is subject to significant model uncertainties. While the fiducial values assumed ($f_{*}=0.1$ and $\varepsilon_0=0.3$) are physically and observationally motivated, lower values may also apply, which motivate our pessimistic case where $f_* \varepsilon_0$ is 1\,dex lower. In particular, efficient mixing of SN ejecta and the ambient ISM can result in more rapid radiative cooling before bubble breakout \citep{Kim_2017,Fielding_2018}, whereas the post-breakout expansion into circumgalactic gas whose density can be anisotropic, inhomogeneous, and generally higher than what we assume \citep{Madau_2001,Scannapieco_2002,Brooks_2009} can also effectively lower $\varepsilon_0$. Meanwhile, the halo mass dependence of model parameters like $f_{*}$ and $f_\mathrm{esc}$ \citep{Khaire_2016,SunFurlanetto_2016,Li_2023,Mutch_2024,SippleLidz_2024} can complicate $y$ signals. Moreover, likely not all haloes have active SNe-driven outflows, though we note the significant longer timescale for inverse Compton cooling ($\sim500$\,Myr) compared to that for burst cycles of star formation ($\sim50$\,Myr) in high-$z$ galaxies \citep{FurlanettoMirocha_2022}. Furthermore, the presence of a more top-heavy IMF and/or Population~III stars may increase the mean energy released per SN by a factor of 2--3 \citep{WoosleyWeaver_1995}. Future work should quantify how these model complexities impact $C^{yy}_{\ell}$ and $C^{y\mathrm{g}}_{\ell}$. As implied by our detectability forecasts, if model uncertainties in e.g. $f_{*}$ and $\omega_\mathrm{SN}$ related to the total budget of SN energy can be narrowed by ancillary galaxy observations, it is promising to constrain $\varepsilon_0$ with $C^{y\mathrm{g}}_{\ell}$ and shed light on the physics of SNe-driven galactic winds at high $z$.

Another simplification made here is the low-$z$ contributions to $C^{y\mathrm{g}}_{\ell}$. Our simplistic model prevents us from physically describing the thermal energy deposited by low-$z$ haloes, especially the massive ones hosting resolved or unresolved galaxy clusters responsible for the majority of the observable tSZ signal. It is possible that the simulation-based predictions we adopt from \citet{Baxter_2021} do not accurately capture the true level of low-$z$ signals or the effectiveness of masking. Building data-driven models across redshift in future work will therefore be helpful. The $y$--galaxy cross-correlation described in Section~\ref{sec:results:cyg} may be attempted at lower redshift with existing data (e.g. Planck/ACT and DESI) for such purposes. 

Finally, some physical factors not considered in our analysis may have useful measurable effects. For example, with sufficiently high angular resolution, distinctions in the scale/redshift dependence of different sources of thermal energy (e.g. galactic outflows versus reionized bubbles) may be utilized for component separation on intermediate scales where non-linear clustering dominates. This is helpful for isolating and exclusively constraining SNe feedback and galactic outflows with $C^{y\mathrm{g}}_{\ell}$. It is thus instructive to extend the current modeling framework and self-consistently predict the size evolution of ionized bubbles during reionization in future studies. Alternatively, one may also stack on galaxies of different types, e.g. starburst versus quiescent galaxies, to narrow down the strength of outflow signals. 

\section{Conclusions}

In summary, we have presented in this Letter a physically motivated model that allows us to calculate and analyze the $y$-type distortion of the CMB spectrum and its anisotropies induced by high-$z$ galaxies, especially their SNe-driven outflows. Our model predicts a relatively large $\bar{y} \sim 10^{-7}$--$10^{-6}$ associated with high-$z$ galaxies, primarily powered by the thermal energy of outflows. While still in good agreement with observational constraints ($5.4\times10^{-8}<\bar{y}<2.2\times10^{-6}$), this higher level of $\bar{y}$ implies large-scale anisotropies of $y$ stronger than many previous models predict. We have demonstrated that, in cross-correlation with forthcoming wide-area surveys of LBGs such as Roman/HLWAS and Rubin/LSST, the planned LiteBIRD mission can potentially measure $y$ anisotropies induced by high-$z$ galactic outflows at high statistical significance up to $z\sim8$. 

\section*{Acknowledgments}

We thank the anonymous reviewer for their helpful and constructive comments. We are indebted to Greg Bryan, Claude-Andr\'{e} Faucher-Gigu\`{e}re, Drummond Fielding, and Natsuko Yamaguchi for helpful discussions. GS was supported by a CIERA Postdoctoral Fellowship. SRF was supported by NASA through award 80NSSC22K0818 and by the National Science Foundation through award AST-2205900. AL acknowledges support from NASA ATP grant 80NSSC20K0497.

\section*{Data Availability}

The data supporting the plots and analysis in this article are available on reasonable request to the corresponding author.



\bibliographystyle{mnras}
\bibliography{tSZ} 



\bsp	
\label{lastpage}
\end{document}